%% file: main.tex
\documentclass[10pt,twocolumn]{article}
\usepackage[margin=0.75in]{geometry}
\usepackage{graphicx}
\usepackage{amsmath}
\usepackage{amssymb}
\usepackage{times}
\usepackage{titlesec}
\usepackage{abstract}
\usepackage{caption}
\usepackage{authblk}
\usepackage{url}
\usepackage{booktabs}
\usepackage{longtable}

\titleformat{\section}{\normalfont\large\bfseries}{\thesection}{1em}{}
\titleformat{\subsection}{\normalfont\bfseries}{\thesubsection}{1em}{}

\title{\bfseries Neckpinches in a Positively Curved Market: Ricci Flow with Surgery on the Nikkei-225}
\author[1]{Siddharta Laloux}
\author[2]{Bhargavi Srinivasan}
\affil[1]{Brown University, Providence, RI 02912, USA}
\affil[2]{CNRS, Laboratoire de Physique Th\'eorique et Mod\`eles Statistiques, Universit\'e Paris-Saclay, 91405 Orsay, France}
\date{}

\begin{document}
\twocolumn[
\begin{@twocolumnfalse}
\maketitle
\begin{abstract}
\noindent
We examine the intrinsic geometry of the Nikkei-225 through the discrete Ollivier-Ricci curvature (ORC) of its empirical correlation graph, using the graph Ricci flow and surgery methodology of Ref.~\cite{srinivasan2025}. The 226-node correlation graph, built from daily returns between January 2020 and June 2026, is \emph{ORC-spherical}: every edge has strictly positive curvature before any flow is applied, close to the equal-weight value $\kappa\approx0.502$. Under a fixed 20-iteration Ricci flow the curvature distribution polarises into a high-curvature bulk and a low-curvature minority, and cutting the low-curvature edges splits the graph into components of 184 and 42 nodes. We then apply the recursive ARCUS procedure (Adaptive Ricci Curvature Unsupervised Surgery), which stops the flow at each node once a curvature gap ratio $G$ exceeds $G_{\min}$ and the surgery set is stable for three iterations. With $G_{\min}=10$ and a minimum cluster size of 15, ARCUS performs 28 surgeries over 13 levels and ends in 29 terminal clusters of 2--14 nodes. The stopping rule fires at every node (after 7--28 iterations, never at the iteration cap), with $G$ between 10.0 and 92.4; the top-level split is again 184/42 ($G=50.9$, 20 iterations). Eight small subgraphs ($N\le27$) stop with $G<15$, so the deeper levels are conditional on $G_{\min}=10$, and 21 of the 28 splits shed a part smaller than the minimum cluster size, giving a deep, unbalanced hierarchy. Every subgraph is ORC-spherical at the start of its flow. The ARCUS top-level split has exactly the same members as the fixed-schedule split, and many terminal clusters are single industries by inspection -- among them nine banks, five insurers, five general trading companies, three steelmakers and three shipping lines. Two stocks delisted during the window, for which no price data could be retrieved, enter as zero-correlation placeholders and fall together in a cluster of food and consumer companies; with them excluded, every terminal cluster is more strongly correlated internally than with the rest of the index.
\end{abstract}
\vspace{0.5cm}
\end{@twocolumnfalse}
]

\section{Introduction}
Stock returns within a market index are, on average, positively correlated, and the resulting correlation matrix has long been used both as a risk-management tool and as the basis for constructing a network representation of the market~\cite{bouchaud2003,mantegna1999,onnela2003}. Because the raw correlation matrix of a broad index is fully connected and its edge weights are closely spaced, most existing network approaches first prune it -- typically to a minimum spanning tree~\cite{mantegna1999,onnela2003} or a planar maximally filtered graph~\cite{tumminello2005} -- before applying community-detection methods~\cite{fortunato2010,newman2006} that implicitly assume a hyperbolic (tree-like) separation between clusters. Pruning of this kind discards information and changes the topology being studied. Applying community detection directly to a correlation matrix is itself delicate, because the usual network null models are not consistent with correlation data~\cite{macmahon2015}.

An alternative, developed in Ref.~\cite{srinivasan2025}, is to keep the full connectivity of the correlation graph and study its intrinsic geometry directly via discrete Ollivier-Ricci curvature~\cite{ollivier2009,linluyau2011}. A weighted graph is called \emph{ORC-spherical} if every edge carries strictly positive curvature before any flow is applied -- the discrete analogue of Hamilton's setting of manifolds of strictly positive Ricci curvature~\cite{hamilton1982}. Stock correlation networks sit naturally in this regime: because most constituents share substantial exposure to a common market factor, pairwise correlations are almost uniformly high, and the community structure is \emph{latent}. Discrete Ricci curvatures have been proposed as indicators of market fragility and systemic risk~\cite{sandhu2016}, including for the Nikkei-225~\cite{samal2021}, and Ollivier-Ricci curvature and Ricci flow have been used for community detection~\cite{ni2019,sia2019}. Unlike sparse, mixed-curvature networks in which inter-community edges already carry negative curvature before any flow, an ORC-spherical graph gives Ricci flow no pre-existing contrast to amplify; any separation has to be generated by the flow itself. The corresponding continuum phenomenon is the neckpinch singularity that develops under Ricci flow on heterogeneous positively curved manifolds~\cite{hamilton1995,angenentknopf2004}, where surgery continues the flow past the singularity~\cite{hamilton1995,perelman2003}. ARCUS is a recursive surgery procedure whose stopping rule and cut are determined from a curvature gap ratio rather than set by the user~\cite{srinivasan2025}.

The Japanese market has been studied with correlation-based methods before: minimum spanning trees have been used to test the conventional industrial grouping of Japanese stocks~\cite{jung2008}, and recursive modularity optimisation to cluster about 1{,}400 of them~\cite{isogai2014}. The Nikkei's decline after 1990 is also a classic case study in econophysics~\cite{johansen1999}. In this paper we study the Nikkei-225 correlation graph with Ricci-flow machinery. We first show that it is ORC-spherical and follow its curvature under a fixed-schedule Ricci flow, followed by a single surgery (Sections~3--4). We then run ARCUS recursively and report the resulting hierarchy, the diagnostics at every node, and the membership of its clusters (Section~5), together with the limitations that currently constrain how the results should be read (Section~6).

\section{Data and Graph Construction}
\label{sec:data}
\textbf{Sample and sampling frequency.} We use daily closing prices from Yahoo!\ Finance, downloaded for 1 January 2010 -- 12 June 2026. The analysis uses only the 1{,}572 Tokyo Stock Exchange trading sessions from 6 January 2020 (the first session of 2020) to 12 June 2026 inclusive, and the earlier prices are not used. From these we compute daily close-to-close simple returns $r_i(t)=P_i(t)/P_i(t-1)-1$ (about 1{,}570 per stock) and the Pearson correlation matrix $C_{ij}$. This gives a complete, undirected, weighted graph on 226 nodes (25{,}425 edges). With $N=226$, $T/N\approx7$, so the empirical correlation matrix carries substantial estimation noise~\cite{laloux1999,plerou2002,bun2017}; we do not filter it, so that the flow acts on the raw matrix, as in Ref.~\cite{srinivasan2025}.

\textbf{Constituent universe.} The node set is a fixed list of Nikkei-225 tickers. Of the 227 tickers in the list, 226 were queried. The list is not updated for changes in the index during the window, so the analysis concerns a fixed cross-section rather than the index as composed on each date, and no survivorship correction is made. The list predates the review of 1 April 2026: it contains GS Yuasa (6674), Casio Computer (6952) and Hino Motors (7205), which left the index on that date, and not Kioxia Holdings, Pan Pacific International Holdings or ARCHION, which joined it. It also contains both NTT Data Group (9613), removed on 4 July 2025, and Rohm (6963), which replaced it. All 226 nodes are listed with their Tokyo Stock Exchange codes in Table~\ref{tab:members} (Appendix~B). The ancillary file \texttt{anc/S1\_constituents.csv} gives, for each node, its ticker, company name, terminal cluster, ARCUS node and any data or index-membership note.

\textbf{Missing data.} Two tickers returned no data because they have been delisted, although both traded for most of the window: NTT Data Group (delisted 26 September 2025 after a tender offer by NTT; 89\% of the window's sessions) and Hino Motors (delisted 30 March 2026 on the formation of ARCHION with Mitsubishi Fuso; 97\%). They are kept as placeholder nodes with $C_{ij}=0$ for all $j\ne i$, which makes each equidistant from every other node. A placeholder does not change any correlation between other stocks, since Pearson correlations are computed pair by pair, but it does enter the graph. Together the two add 449 of the 25{,}425 edges, all at the maximal distance, and they appear in every node's neighbour measure. Because the neighbour weights are $\propto e^{-w_{ij}^2}$ (Section~3), the two together carry at most 0.8\% of any real node's non-lazy mass, under either scaling. No other stock is zero-filled. Three stocks were listed during the window: Nippon Express Holdings (9147, listed 4 January 2022; 69\% of sessions), Shizuoka Financial Group (5831, 3 October 2022; 57\%) and Socionext (6526, 12 October 2022; 57\%). Their correlations use the dates on which both stocks trade (pairwise-complete observations). Because different pairs then use different samples, $C$ is not positive semidefinite: it has one negative eigenvalue, $-0.097$, whose eigenvector is concentrated on Shizuoka Financial Group and Socionext.

\textbf{Edge weights.} Edge weights are correlation distances. We write $w_{ij}^{(0)}$ for the initial weight and $w_{ij}^{(l)}$ for the weight after $l$ Ricci-flow iterations (Section~3); $d_w(i,j)$ is the shortest-path distance under the current weights. Two scalings of $w^{(0)}$ are used:
\begin{equation}
w_{ij}^{(0)}=
\begin{cases}
\sqrt{1-C_{ij}} & \text{(Section 4)},\\
\sqrt{2(1-C_{ij})} & \text{(Section 5)},
\end{cases}
\label{eq:weights}
\end{equation}
the second as in Ref.~\cite{srinivasan2025}. They differ by a factor of $\sqrt2$. Although $C$ is not positive semidefinite, we checked directly that both satisfy the triangle inequality for every triple of nodes, so initially $d_w(i,j)=w_{ij}^{(0)}$. The scaling matters because, in the \texttt{GraphRicciCurvature} implementation~\cite{ni2019code} that we use, the lazy random-walk measure assigns each neighbour a share of the non-lazy mass proportional to $\exp(-w_{ij}^2)$, so curvature is \emph{not} invariant under a uniform rescaling of the weights. The two sections should therefore be read as separate runs; for example, the minimum root curvature is $0.468$ in Section~4 and $0.4338$ in Section~5. Both runs use the same correlation matrix: recomputing the initial curvatures stored by the fixed-schedule run from this matrix reproduces them to within $10^{-15}$, and the ARCUS run reads the same file.

\section{Ollivier-Ricci Curvature and Ricci Flow}
For a weighted graph $G=(V,E,w)$, the Ollivier-Ricci curvature of an edge $ij$ is
\begin{equation}
\kappa_{ij} = 1 - \frac{W(m_i, m_j)}{d_w(i,j)},
\end{equation}
where $d_w(i,j)$ is the shortest-path distance and $W$ is the Wasserstein (optimal transport) distance between the lazy random-walk measures $m_i, m_j$ centred on nodes $i$ and $j$, with idleness parameter $\alpha=1/2$~\cite{ollivier2009,linluyau2011}.

\textbf{Definition (ORC-spherical graph, following Ref.~\cite{srinivasan2025}).} A weighted graph $G=(V,E,w)$ is ORC-spherical if $\kappa_{ij} > 0$ for every edge $(i,j) \in E$. An edge is approximately flat if $\kappa_{ij}\approx 0$, and hyperbolic if $\kappa_{ij}<0$.

For a fully connected graph on $n$ nodes with equal edge weights, all edges share the curvature
\begin{equation}
\kappa = (1-\alpha)\,\frac{n}{n-1},
\end{equation}
which at $\alpha=1/2$ and $n=226$ gives $\kappa \approx 0.5022$ -- the maximally degenerate ORC-spherical case, and a natural reference point for the empirical graph. (With equal weights the neighbour measure is uniform, so this value does not depend on the weight scale.)

The discrete Ricci flow updates each edge weight by its curvature and then rescales all weights so that their mean is one, which prevents collapse:
\begin{equation}
\begin{gathered}
\tilde w_{ij}^{(l+1)} = \bigl(1-\kappa_{ij}^{(l)}\bigr)\, w_{ij}^{(l)},\\
w_{ij}^{(l+1)} = \tilde w_{ij}^{(l+1)}\,\frac{|E|}{\sum_{(k,m)\in E}\tilde w_{km}^{(l+1)}}.
\end{gathered}
\label{eq:flow}
\end{equation}
The update contracts positively curved (intra-community) edges and stretches near-flat and hyperbolic (inter-community) edges~\cite{ni2019}. This is the update of \texttt{GraphRicciCurvature} (version~0.6.1, step size 1), and it is used in both the fixed-schedule and the ARCUS runs below. We use $\alpha=1/2$ and the OTD (optimal transport distance) method throughout.

\section{Results: Fixed-Schedule Flow and a Single Surgery}

\subsection{Initial curvature confirms an ORC-spherical graph}
\begin{figure}[h]
\centering
\includegraphics[width=0.85\linewidth]{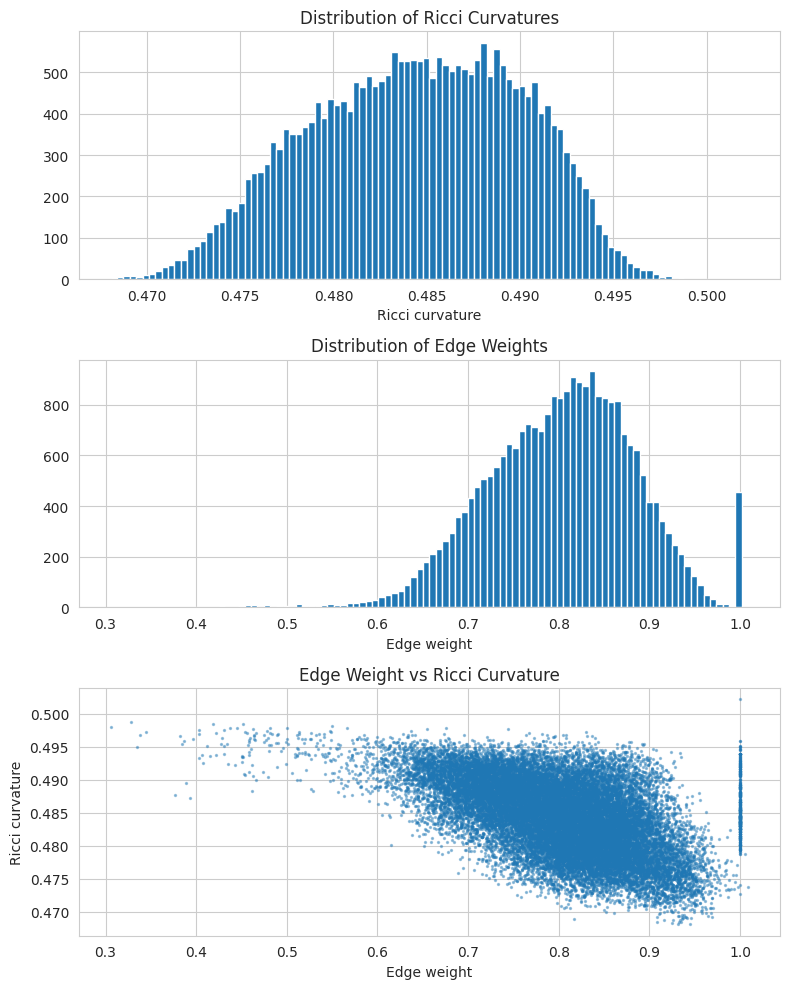}
\caption{Nikkei-225 correlation graph before Ricci flow (0 iterations; $w^{(0)}=\sqrt{1-C}$, the Section-4 scaling of Eq.~\eqref{eq:weights}): Ricci curvature histogram (top), edge-weight histogram (middle), and edge weight vs.\ curvature scatter (bottom). All curvatures are strictly positive.}
\label{fig:initial}
\end{figure}

Figure~\ref{fig:initial} shows the curvature and edge-weight distributions of the unflowed Nikkei-225 graph. Every one of the 25{,}425 edges has strictly positive curvature, concentrated between 0.468 and 0.502. By the Definition above the graph is therefore ORC-spherical, sitting just below the equal-weight value of 0.5022. There is no pre-existing bimodal structure and no gap between potential intra- and inter-community links at this stage.

\subsection{Evolution under Ricci flow}
\begin{figure*}[t]
\centering
\includegraphics[width=0.36\textwidth]{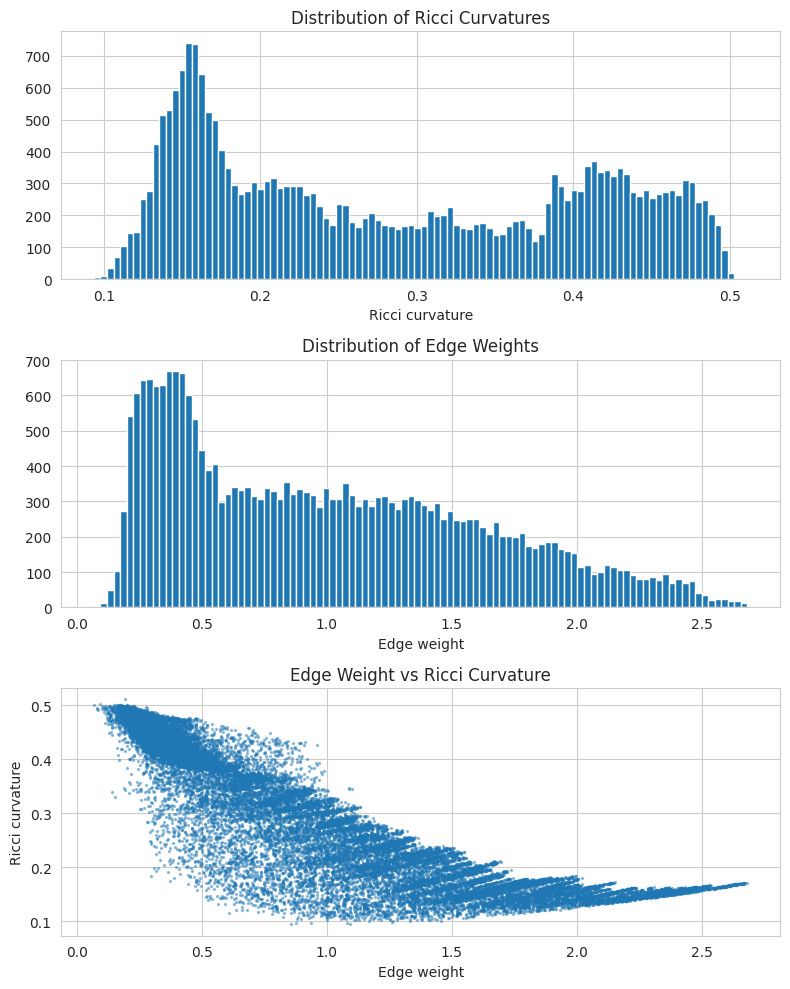}\hspace{0.06\textwidth}
\includegraphics[width=0.36\textwidth]{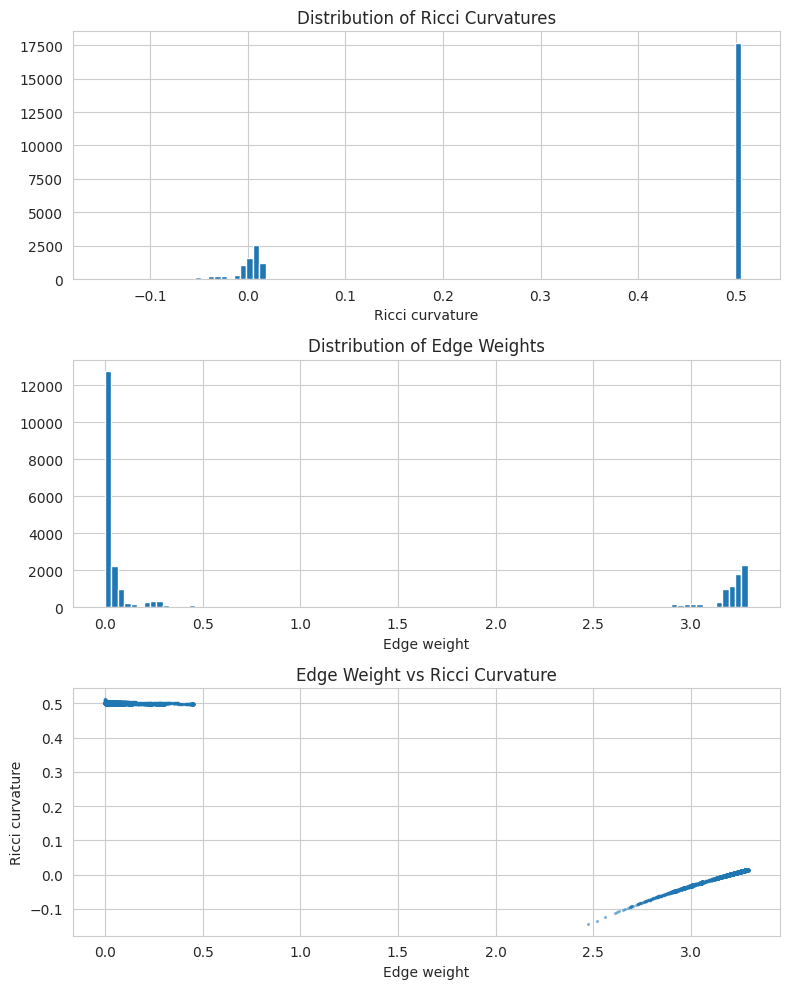}
\caption{Curvature/weight distributions (curvature histogram, edge-weight histogram, edge weight vs.\ curvature) after 10 Ricci flow iterations (left) and after 20 iterations (right).}
\label{fig:evolution}
\end{figure*}

Table~\ref{tab:evolution} follows five edges, identified by stock pair, through checkpoints at 0, 1, 5, 10, 15 and 20 iterations. Three lie within groups that stay together after surgery (two megabanks, two semiconductor-equipment makers and two automakers). The other two cross the eventual 184/42 cut: Toyota--Nintendo, and Takeda--Daiichi Sankyo, two pharmaceutical companies that end up on opposite sides. Over the first ten iterations the distribution broadens steadily (Figure~\ref{fig:evolution}, left): the tight peak near 0.49 spreads out to between 0.09 and 0.51, with a second population forming near $\kappa\approx0.15$. The two cross-cut edges fall fastest. Between 12 and 19 iterations the flow passes through a transient. Up to 3{,}273 edges have negative curvature (at iteration 17), and some edges inside groups dip before recovering, as Tokyo Electron--Advantest does at iteration 15: 1{,}957 edges are below 0.25 at iteration 15 but above it at iteration 20. By iteration 19 the low population has settled at 7{,}728 edges. At 20 iterations the distribution is sharply bimodal (Figure~\ref{fig:evolution}, right): 17{,}697 edges lie between 0.497 and 0.512 and 7{,}728 between $-0.146$ and $0.014$, with none in between -- the neckpinch signature. The flowed weights separate in the same way.

\begin{table}[h]
\centering
\small
\setlength{\tabcolsep}{4pt}
\begin{tabular}{rrrrrr}
\toprule
 & \multicolumn{3}{c}{within groups} & \multicolumn{2}{c}{across the cut} \\
\cmidrule(lr){2-4}\cmidrule(lr){5-6}
Iter. & MUFG & TEL & Toyota & Toyota & Takeda \\
 & SMFG & Advantest & Honda & Nintendo & Daiichi S. \\
\midrule
0  & 0.498 & 0.497 & 0.494 & 0.481 & 0.491 \\
1  & 0.496 & 0.495 & 0.489 & 0.462 & 0.480 \\
5  & 0.498 & 0.496 & 0.473 & 0.365 & 0.413 \\
10 & 0.500 & 0.472 & 0.433 & 0.138 & 0.103 \\
15 & 0.499 & $-0.239$ & 0.465 & 0.059 & $-0.006$ \\
20 & 0.502 & 0.499 & 0.500 & 0.011 & 0.005 \\
\bottomrule
\end{tabular}
\caption{Curvature $\kappa_{ij}^{(l)}$ of five fixed edges, identified by stock pair, during the fixed-schedule flow (Section-4 scaling). MUFG: Mitsubishi UFJ Financial Group (8306); SMFG: Sumitomo Mitsui Financial Group (8316); TEL: Tokyo Electron (8035); Advantest (6857); Toyota Motor (7203); Honda Motor (7267); Nintendo (7974); Takeda Pharmaceutical (4502); Daiichi Sankyo (4568).}
\label{tab:evolution}
\end{table}

\subsection{Emergent network structure and a single surgery}
\begin{figure*}[t]
\centering
\includegraphics[width=0.47\textwidth]{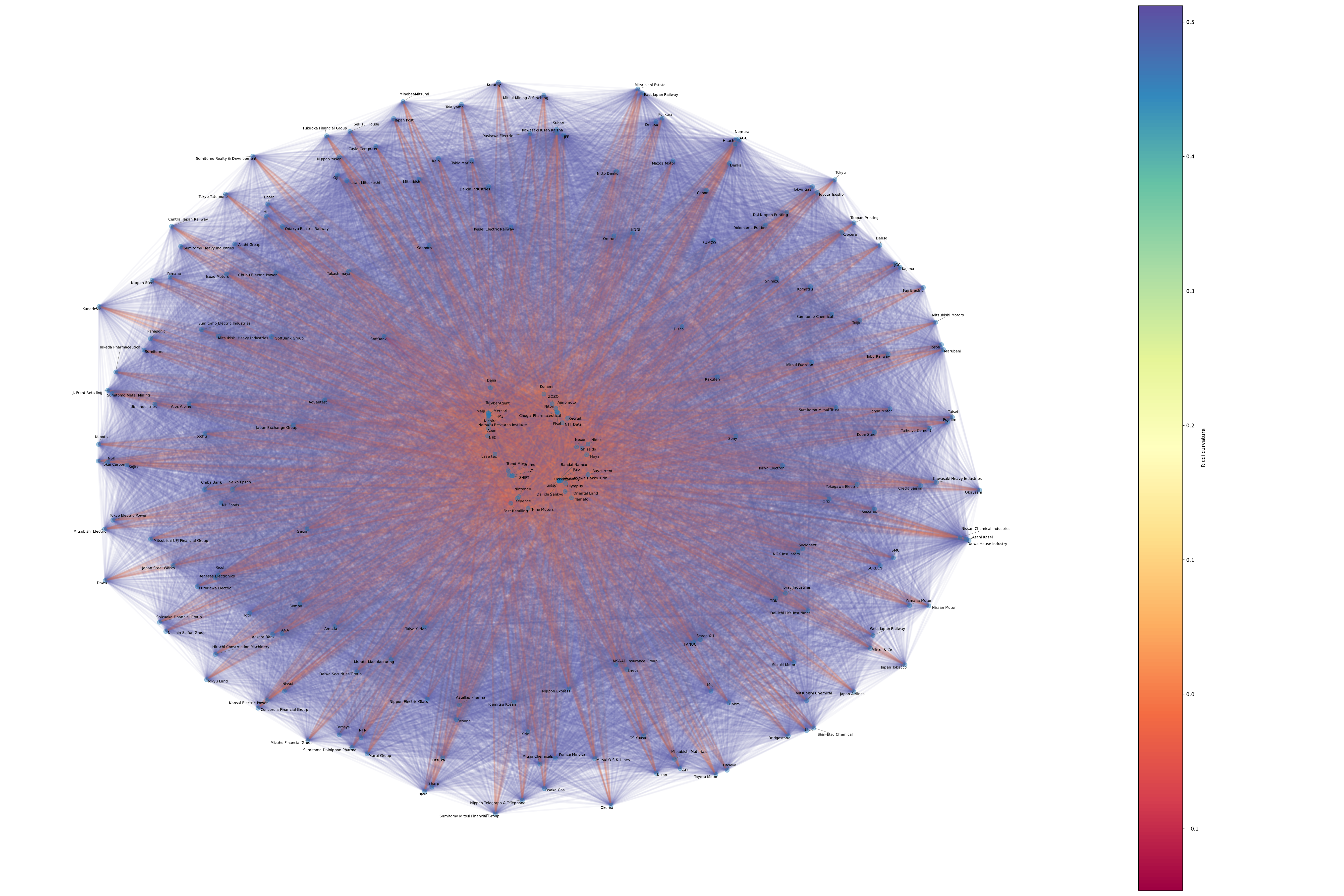}\hfill
\includegraphics[width=0.47\textwidth]{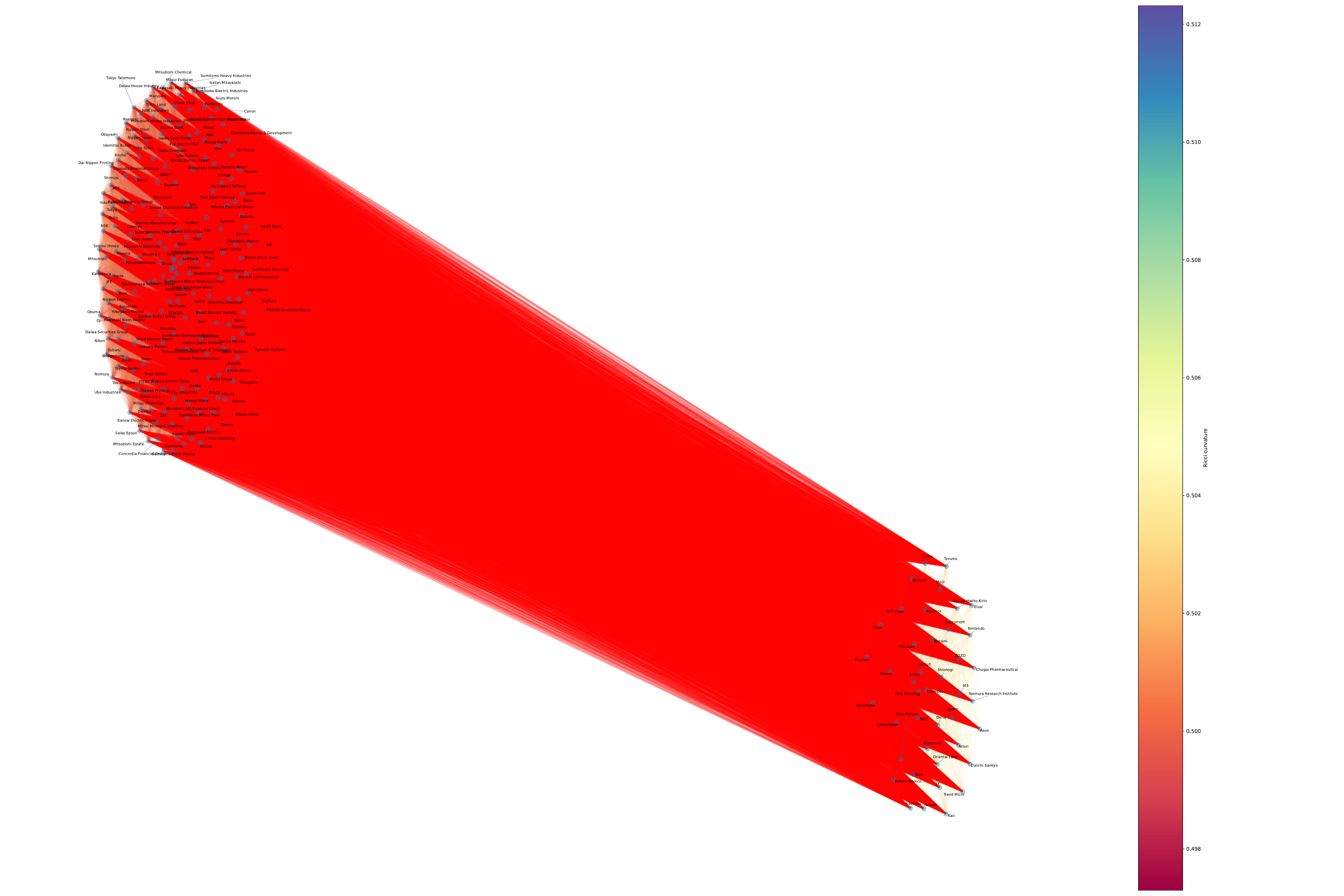}
\caption{(a) Spring-layout embedding of the Nikkei-225 graph after 20 Ricci flow iterations, edges coloured by Ricci curvature. (b) Cut edges (red) identified by the surgery of Eq.~\eqref{eq:cut} on the 20-iteration graph, connecting the 184-node and 42-node components. Nodes are labelled by company short name; the labels are vector text, so the panels can be enlarged in the PDF.}
\label{fig:netcut}
\end{figure*}

\begin{figure}[h]
\centering
\includegraphics[width=0.85\linewidth]{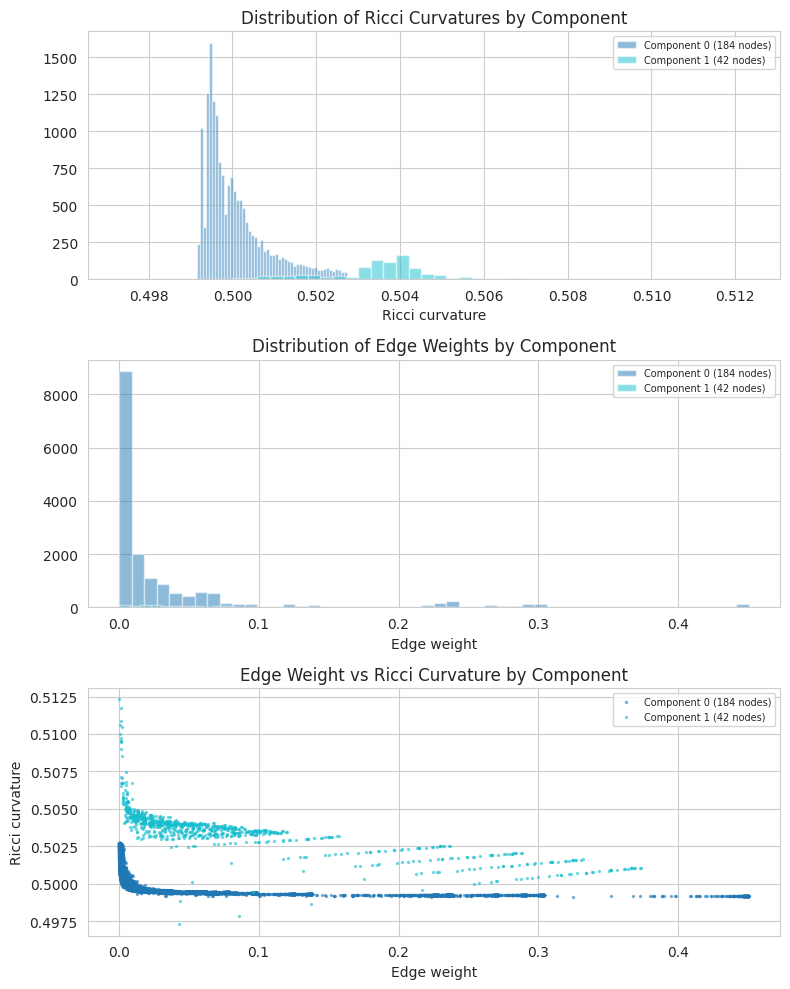}
\caption{Curvature and edge-weight distributions of the two post-surgery components, computed on the initial weights $w^{(0)}=\sqrt{1-C}$.}
\label{fig:components}
\end{figure}

Figure~\ref{fig:netcut}a shows the 20-iteration graph, coloured by edge curvature: a clear two-lobe, dumbbell-like structure is visible. Line surgery removes every edge whose curvature after the flow is small relative to its flowed weight,
\begin{equation}
\kappa_{ij}^{(20)} < 0.2\, w_{ij}^{(20)},
\label{eq:cut}
\end{equation}
where $w^{(20)}$ is the flowed weight, normalised to mean one (Eq.~\eqref{eq:flow}); the factor 0.2 was chosen by inspection. The cut removes 7{,}728 of 25{,}425 edges and splits the graph into two connected components of 184 and 42 nodes (Figure~\ref{fig:netcut}b and Figure~\ref{fig:components}). The result does not depend on the details of this rule. The same 7{,}728 edges are removed if $w^{(0)}$ is used in place of $w^{(20)}$, and they are exactly the edges below the empty curvature gap $[0.014,0.497]$ at iteration 20, so any threshold inside that gap gives the same surgery. The modularity~\cite{newman2006} of the 184/42 partition, computed on the post-surgery graph with the flowed weights $w^{(20)}$ as edge weights, is 0.183.

\section{Results: Recursive ARCUS Surgery}
\label{sec:arcus}

\subsection{The procedure as implemented}
ARCUS replaces the fixed schedule and the hand-chosen threshold of Eq.~\eqref{eq:cut} with a data-driven stopping rule and a recursion~\cite{srinivasan2025}. At each node of the hierarchy, our implementation proceeds as follows.
\begin{enumerate}
\item Rebuild the complete weighted graph on the node's stocks from the \emph{original} correlation submatrix, with the Section-5 scaling $w^{(0)}_{ij}=\sqrt{2(1-C_{ij})}$ of Eq.~\eqref{eq:weights}, and compute the initial curvatures (recording whether all are positive).
\item Run the Ricci flow of Eq.~\eqref{eq:flow} one iteration at a time. After each iteration, sort the edge curvatures, locate the \emph{largest gap} between consecutive values, $[\kappa_{\text{lo}},\kappa_{\text{hi}}]$, and split the edges into the populations below and above it. Compute the curvature gap ratio
\begin{equation}
G = \frac{\mu(\kappa^{+}) - \mu(\kappa^{-})}{\sigma_{\text{pooled}}},
\end{equation}
where $\kappa^{+}$ and $\kappa^{-}$ are the upper and lower populations and $\sigma_{\text{pooled}}=\bigl[\tfrac12\bigl(\sigma(\kappa^{+})^2+\sigma(\kappa^{-})^2\bigr)\bigr]^{1/2}$. This is an unsupervised analogue of the curvature gap of Gosztolai and Arnaudon~\cite{gosztolai2021}: $G\gg1$ means the two populations are sharply separated, so any cut placed in the empty gap gives the same surgery set.
\item Stop once at least 5 iterations have run, $G>G_{\min}$, and the number of edges in the lower population has been identical for three consecutive iterations (hard cap: 100 iterations).
\item Perform surgery by removing every edge with $\kappa\le\kappa_{\text{lo}}$ -- an absolute curvature threshold, unlike Eq.~\eqref{eq:cut}. Return the largest connected component as part A and the union of all remaining nodes as part B.
\item Recurse on any part with at least $N_{\min}$ nodes; smaller parts are terminal.
\end{enumerate}
The parameters are $\alpha=1/2$, $G_{\min}=10$, $N_{\min}=15$, a stability window of 3 iterations, and a minimum of 5 iterations. Note that $G_{\min}=10$ here, whereas $G_{\min}=15$ was used for the NASDAQ-100 in Ref.~\cite{srinivasan2025}.

\subsection{Hierarchy and per-node diagnostics}
\begin{figure*}[t]
\centering
\includegraphics[width=\textwidth]{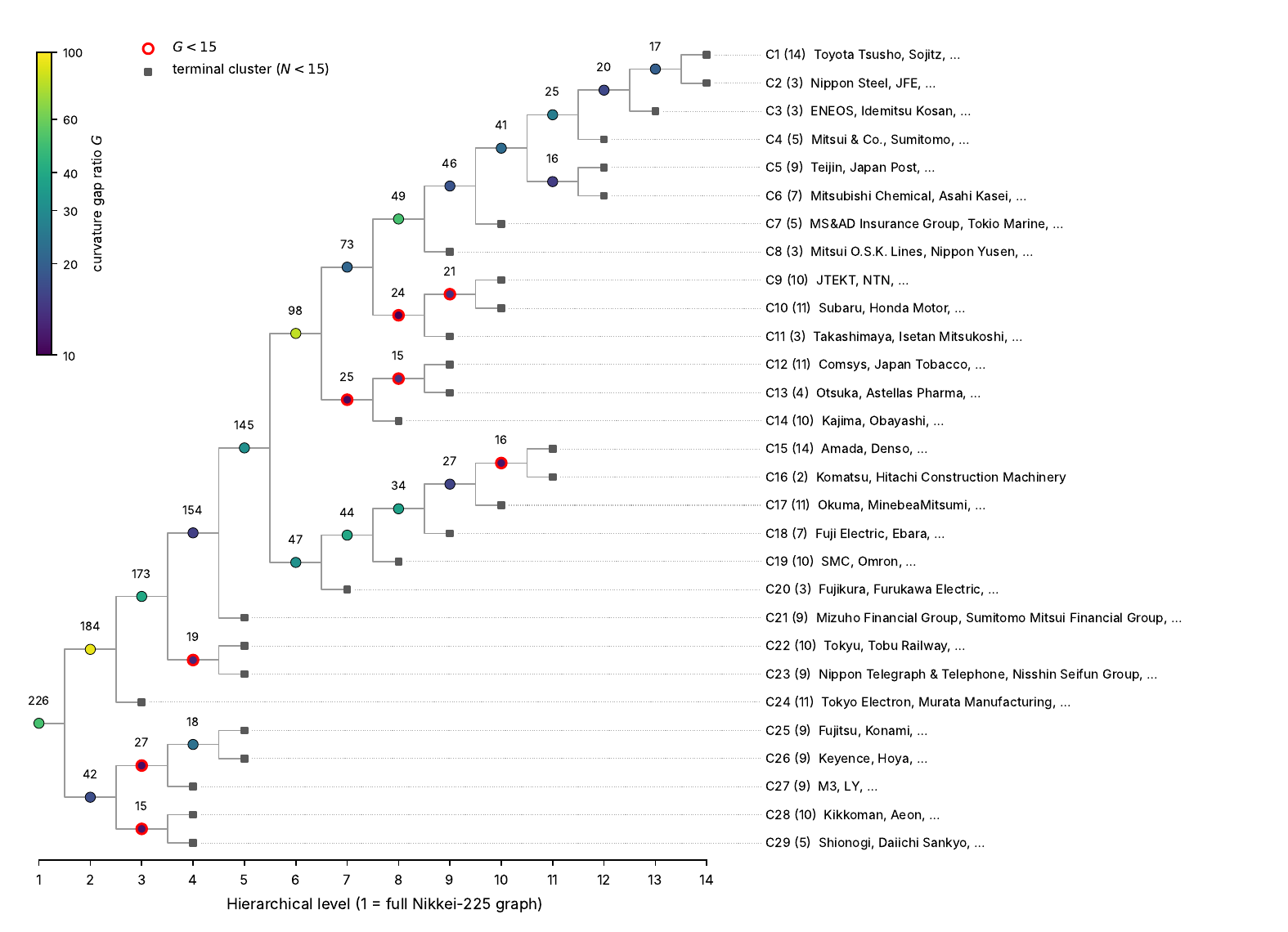}
\caption{ARCUS hierarchy for the Nikkei-225 (28 surgeries, 29 terminal clusters). Circles are surgery nodes, labelled with $N$ and coloured by the curvature gap ratio $G$ at stopping; red rings mark $G<15$. Squares are terminal clusters ($N<15$). Part A (the largest component) is drawn above part B at each split. Terminal clusters are labelled with an identifier, their size and their two most central members (highest mean correlation with the rest of the cluster); full membership is in Table~\ref{tab:members}.}
\label{fig:hier}
\end{figure*}

Table~\ref{tab:arcus} gives the first three levels, and Figure~\ref{fig:hier} the full hierarchy; the complete per-node diagnostics for all 28 surgeries are in Table~\ref{tab:full} and the membership of all 29 terminal clusters in Table~\ref{tab:members} (Appendix).

\textbf{Level 1.} The root graph (226 nodes) is ORC-spherical with minimum curvature $\kappa_{\min}=0.4338$. The stopping rule fires after 20 iterations with $G=50.93$, the largest curvature gap lying between $\kappa=0.0116$ and $\kappa=0.4993$. Surgery removes 7{,}728 edges, which is exactly $184\times42$, the number of edges between the two parts, and yields components of 184 and 42 nodes. This is the same pair of component sizes found by the fixed-schedule surgery of Section~4 -- which also used 20 iterations, but a different weight scaling and the manually chosen threshold of Eq.~\eqref{eq:cut}.

\textbf{Level 2.} The 184-node component splits 173/11 ($G=92.44$ after 22 iterations, the largest $G$ in the hierarchy) and the 42-node component splits 27/15 ($G=17.19$ after 16 iterations).

\textbf{Whole hierarchy.} With $N_{\min}=15$ the recursion performs 28 surgeries over 13 levels and ends in 29 terminal clusters containing 2 to 14 nodes (total 226). The stopping rule fires through the $G$-and-stability criterion at all 28 nodes; none stops at the iteration cap. Stopping iterations range from 7 to 28 (median 17), and $G$ ranges from 10.02 to 92.44 (median 18.97). The gap interval $[\kappa_{\text{lo}},\kappa_{\text{hi}}]$ has $\kappa_{\text{lo}}\in[0.0055,0.1120]$ and $\kappa_{\text{hi}}\in[0.2612,0.5275]$, so any single threshold in $(0.112,0.261)$ lies inside the empty gap at all 28 nodes.

\textbf{Surgery sets.} At 26 of the 28 nodes the surgery set is exactly the set of edges between the two parts ($n_{\text{surgery}}=n_A n_B$). At the other two, additional low-curvature edges were also removed: 3{,}010 edges against 2{,}926 at node \texttt{1.1.1}, and 120 against 110 at node \texttt{1.1.1.1.1.1.1.2.1}. Rerunning node \texttt{1.1.1.1.1.1.1.2.1} from the correlation matrix reproduces the recorded $G=13.70$, gap $[0.0807,0.3447]$ and 120-edge surgery set. The surgery leaves three components: ten stocks that form part A (C9); nine vehicle makers with Bridgestone; and Yokohama Rubber on its own. The two ten-stock components tie for largest. The ten extra edges are those joining Yokohama Rubber to the other ten stocks of part B, so C10 is not a single connected group. For node \texttt{1.1.1} the saved output does not determine whether part B is connected.

\textbf{Shape.} The hierarchy is deep and unbalanced: in 21 of 28 splits the smaller part has fewer than $N_{\min}=15$ nodes and is terminal immediately, and only 7 splits produce two parts that are both recursed into. No node with $N\ge N_{\min}$ failed to split, so every branch ends because of the size cut-off rather than because surgery stopped finding a gap.

\begin{table}[h]
\centering
\small
\begin{tabular}{llcccc}
\toprule
Level & Node & $N$ & Surgery links & $G$ (iter) & A/B \\
\midrule
1 & \texttt{1}   & 226 & 7{,}728 & 50.93 (20) & 184/42 \\
2 & \texttt{1.1} & 184 & 1{,}903 & 92.44 (22) & 173/11 \\
2 & \texttt{1.2} &  42 & 405     & 17.19 (16) & 27/15  \\
3 & \texttt{1.1.1} & 173 & 3{,}010 & 39.66 (20) & 154/19 \\
3 & \texttt{1.2.1} &  27 & 162     & 10.85 (12) & 18/9   \\
3 & \texttt{1.2.2} &  15 & 50      & 11.35 (9)  & 10/5   \\
\bottomrule
\end{tabular}
\caption{ARCUS for the Nikkei-225, levels 1--3 ($\alpha=0.5$, $G_{\min}=10$, $N_{\min}=15$). The iteration at which the stopping rule fired is in parentheses; A/B are the sizes of the two parts. Level~3 omits node \texttt{1.1.2} (11 nodes, terminal). Full table in Appendix.}
\label{tab:arcus}
\end{table}

\subsection{Cluster membership}
\label{sec:members}
\textbf{Level 1.} The ARCUS top-level split has exactly the same members as the fixed-schedule surgery of Section~4. The 42-node part (clusters C25--C29) contains pharmaceutical companies (Chugai, Daiichi Sankyo, Eisai, Kyowa Kirin, Shionogi), food and consumer companies (e.g.\ Ajinomoto, Kikkoman, Meiji, Kao), games, internet and IT-services companies (e.g.\ Nintendo, Konami, Nexon, Mercari, CyberAgent, Fujitsu, NEC), precision and medical-equipment makers (e.g.\ Hoya, Olympus, Terumo, Keyence), and the two placeholder nodes. Its members are less correlated with the rest of the index than those of the 184-node part: the mean correlation of a stock with all others is 0.256 against 0.377 (placeholders excluded).

\textbf{Main branch.} The first groups to separate from the 184-node part are recognisable industries. At Level~2 eleven semiconductor and electronic-component makers split off (C24: Advantest, Disco, Tokyo Electron, SCREEN, SUMCO, Renesas, Rohm, Murata, TDK, Taiyo Yuden and Yaskawa Electric); a semiconductor group also separates at Level~2 in the NASDAQ-100~\cite{srinivasan2025}. At Level~3 a 19-stock group splits off and divides into eight railway companies with ANA and Marui Group (C22) and nine utilities, telecoms and food and beverage companies (C23). At Level~4 nine banks separate (C21: the three megabanks, Resona, Sumitomo Mitsui Trust and four regional banking groups).

\textbf{Terminal clusters.} Many terminal clusters are single industries: five insurers (C7), the general trading companies Itochu, Marubeni, Mitsubishi, Mitsui \& Co.\ and Sumitomo (C4), three steelmakers (C2), three oil companies (C3), three shipping lines (C8), three department-store groups (C11), automakers with two tyre makers (C10), builders and property developers (C14), and chemical companies (C6). C10 is not a single connected group: at its parent node the surgery leaves the nine vehicle makers and Bridgestone connected, and Yokohama Rubber on its own (Section~5.2). Pharmaceutical companies form two clusters, one on each side of the Level-1 split (C13 and C29). Other clusters are mixed, for example C12 (utilities, food and drink, tobacco, printing, retail and services) and C5. These industry descriptions are ours, by inspection. This is consistent with the observation that the correlation eigenvectors outside the random-matrix bulk pick out business sectors~\cite{plerou2002}.

\textbf{Cohesion.} For every terminal cluster, the mean internal correlation $\rho_{\text{in}}$ exceeds the mean correlation with the rest of the index $\rho_{\text{out}}$ (Table~\ref{tab:members}, placeholders excluded). The most cohesive are the shipping lines (0.85), banks (0.82), steelmakers (0.79), trading companies (0.77) and department stores (0.77). Against 2{,}000 random groups of the same size, 28 clusters have a larger $\rho_{\text{in}}-\rho_{\text{out}}$ than at least 95\% of them; the exception is C12, the mixed group (12\% of random groups have a larger difference). Because the clusters were obtained from the same correlations, this is a descriptive check of cohesion, not an independent validation.

\textbf{Placeholder nodes.} The two delisted tickers without price data (Hino Motors and NTT Data) fall together in C28, with eight food, consumer and leisure companies (Aeon, Ajinomoto, Kao, Kikkoman, Meiji, Nichirei, Oriental Land and Toho). A placeholder has zero correlation with every stock, so its position reflects missing data rather than market structure.

\subsection{Sensitivity to $G_{\min}$ and sphericity at depth}
Eight of the 28 nodes stop with $G<15$ (Figure~\ref{fig:hier}, red rings): nodes \texttt{1.1.1.1.1.1.1.2} ($G=10.02$), \texttt{1.1.1.1.1.1.2} (10.68), \texttt{1.2.1} (10.85), \texttt{1.1.1.1.1.2.1.1.1.1} (11.34), \texttt{1.2.2} (11.35), \texttt{1.1.1.2} (12.73), \texttt{1.1.1.1.1.1.1.2.1} (13.70) and \texttt{1.1.1.1.1.1.2.1} (14.19). All eight are small ($N\le27$); every node with $N>27$ has $G>15$. Under the NASDAQ-100 setting $G_{\min}=15$ the flow would not have stopped at these nodes at the recorded iteration. Whether continuing the flow changes their splits has not been tested, so the lower levels of the hierarchy are conditional on $G_{\min}=10$.

All 28 subgraphs were ORC-spherical at the start of their own flow: the implementation's per-node flag is true at every node. The minimum curvature of each subgraph was printed during the run but not stored.

\section{Discussion and Limitations}
\textbf{What the results show.} The Nikkei-225 correlation graph is ORC-spherical. Under Ricci flow a curvature gap opens at every one of the 28 nodes that ARCUS visits: the stability-based stopping rule fires each time, $G$ exceeds $G_{\min}$, and at 26 nodes the surgery set is exactly the cut between the two parts. In that sense the Nikkei-225 shows the behaviour that ARCUS is designed to detect. Every subgraph is ORC-spherical, so the same procedure applies at each level. The top-level partition is identical to the one found by the fixed-schedule surgery of Section~4, despite a different weight scaling and cut criterion, and the recursion separates groups that are, by inspection, largely single industries (Section~\ref{sec:members}).

\textbf{Differences from the NASDAQ-100 setting.} Relative to the NASDAQ-100 analysis in Ref.~\cite{srinivasan2025}, the settings differ ($G_{\min}=10$ vs.\ 15) and so does the shape of the output: 13 levels here against five there, with most splits peeling off a small group, often a single industry, from a large core. The universe is also larger (226 vs.\ 101 nodes). We do not attribute the difference in tree shape to market structure, and with eight nodes close to the threshold we cannot say how much of the depth is a consequence of $G_{\min}$.

\textbf{Limitations.}
\begin{itemize}
\item \emph{Descriptive industry reading.} The industry descriptions in Section~\ref{sec:members} are ours, by inspection. The clusters have not been compared with an official sector classification (e.g.\ the Nikkei or TOPIX sector schemes), the test applied to minimum spanning trees of Japanese stocks in Ref.~\cite{jung2008}, and the cohesion statistics use the same correlations that produced the clusters.
\item \emph{$G_{\min}$ and other parameters.} Results are for a single setting ($\alpha=0.5$, $G_{\min}=10$, $N_{\min}=15$). Sensitivity to these, and to the estimation window, has not been examined.
\item \emph{Estimation noise.} With $T/N\approx7$ the correlation matrix is not cleaned~\cite{laloux1999,bun2017}. Whether the hierarchy survives random-matrix filtering, or a null model suited to correlation matrices~\cite{macmahon2015}, has not been tested.
\item \emph{Two weight scalings.} Sections~4 and~5 use different scalings of Eq.~\eqref{eq:weights} and, because curvature depends on the weight scale in this implementation, are not strictly comparable. A like-for-like rerun of Section~4 under the Section-5 scaling would remove this caveat.
\item \emph{Fixed universe and missing data.} The node set is a fixed list that is not updated for index changes (Section~\ref{sec:data}). The two delisted tickers enter the graph as nodes with zero correlation to everything and end up together in C28; their placement reflects missing data rather than market structure, and a rerun without them, or with their prices from another source, would show whether C28 and its neighbours change. The three stocks listed during the window are correlated over shorter samples than the rest.
\item \emph{Binary splits by construction.} Part B is the union of all non-largest components, so ``two parts'' reflects the implementation rather than a measured number of communities.
\end{itemize}

\section{Conclusion}
The Nikkei-225 correlation graph is ORC-spherical, and discrete Ricci flow develops a bimodal curvature distribution whose low-curvature population can be cut to separate the graph into 184- and 42-node components -- first with a fixed 20-iteration schedule and the hand-chosen threshold of Eq.~\eqref{eq:cut}, and again by the adaptive ARCUS criterion. Applied recursively with $G_{\min}=10$ and $N_{\min}=15$, ARCUS stops on its gap-and-stability criterion at all 28 nodes it visits and produces a 13-level hierarchy of 29 terminal clusters, though the lower levels sit close to the $G_{\min}$ threshold and the hierarchy mostly sheds small groups from a large core. Every subgraph is ORC-spherical, and the hierarchy separates recognisable industry groups -- banks, insurers, trading companies, steel, shipping, autos, semiconductors and pharmaceuticals -- while a few clusters are mixed. Comparing the clusters with an official sector classification, rerunning without the two placeholder nodes, and repeating the run at $G_{\min}=15$ are the natural next steps.

\clearpage
\onecolumn
\appendix
\section{Full ARCUS per-node diagnostics}
Table~\ref{tab:full} lists the diagnostics recorded at each of the 28 surgery nodes.

\input{tables/tab_full.tex}

\section{Terminal-cluster membership}
\label{app:members}
\input{tables/tab_members.tex}

\end{document}

%% file: tables/tab_full.tex
\begin{table}[!ht]
\centering
\footnotesize
\begin{tabular}{lrrrrrrrr}
\toprule
Node & Level & $N$ & A/B & Surgery links & $G$ & Iter. & $\kappa_{\text{lo}}$ & $\kappa_{\text{hi}}$ \\
\midrule
\texttt{1} & 1 & 226 & 184/42 & 7{,}728 & 50.93 & 20 & 0.0116 & 0.4993 \\
\texttt{1.1} & 2 & 184 & 173/11 & 1{,}903 & 92.44 & 22 & 0.0239 & 0.4930 \\
\texttt{1.1.1} & 3 & 173 & 154/19 & 3{,}010 & 39.66 & 20 & 0.0476 & 0.5030 \\
\texttt{1.1.1.1} & 4 & 154 & 145/9 & 1{,}305 & 15.29 & 15 & 0.0331 & 0.3506 \\
\texttt{1.1.1.1.1} & 5 & 145 & 98/47 & 4{,}606 & 31.56 & 28 & 0.0121 & 0.3817 \\
\texttt{1.1.1.1.1.1} & 6 & 98 & 73/25 & 1{,}825 & 80.44 & 21 & 0.0086 & 0.4867 \\
\texttt{1.1.1.1.1.1.1} & 7 & 73 & 49/24 & 1{,}176 & 21.27 & 20 & 0.0219 & 0.3893 \\
\texttt{1.1.1.1.1.1.1.1} & 8 & 49 & 46/3 & 138 & 50.74 & 17 & 0.0080 & 0.4984 \\
\texttt{1.1.1.1.1.1.1.1.1} & 9 & 46 & 41/5 & 205 & 18.12 & 14 & 0.0311 & 0.4392 \\
\texttt{1.1.1.1.1.1.1.1.1.1} & 10 & 41 & 25/16 & 400 & 22.15 & 22 & 0.0240 & 0.3618 \\
\texttt{1.1.1.1.1.1.1.1.1.1.1} & 11 & 25 & 20/5 & 100 & 26.38 & 17 & 0.0159 & 0.5247 \\
\texttt{1.1.1.1.1.1.1.1.1.1.1.1} & 12 & 20 & 17/3 & 51 & 16.50 & 13 & 0.0301 & 0.4985 \\
\texttt{1.1.1.1.1.1.1.1.1.1.1.1.1} & 13 & 17 & 14/3 & 42 & 19.82 & 12 & 0.0172 & 0.5275 \\
\texttt{1.1.1.1.1.1.1.1.1.1.2} & 11 & 16 & 9/7 & 63 & 15.06 & 22 & 0.0525 & 0.3422 \\
\texttt{1.1.1.1.1.1.1.2} & 8 & 24 & 21/3 & 63 & 10.02 & 7 & 0.0642 & 0.3737 \\
\texttt{1.1.1.1.1.1.1.2.1} & 9 & 21 & 10/11 & 120 & 13.70 & 18 & 0.0807 & 0.3447 \\
\texttt{1.1.1.1.1.1.2} & 7 & 25 & 15/10 & 150 & 10.68 & 13 & 0.0698 & 0.3270 \\
\texttt{1.1.1.1.1.1.2.1} & 8 & 15 & 11/4 & 44 & 14.19 & 14 & 0.1120 & 0.3702 \\
\texttt{1.1.1.1.1.2} & 6 & 47 & 44/3 & 132 & 31.80 & 17 & 0.0083 & 0.4769 \\
\texttt{1.1.1.1.1.2.1} & 7 & 44 & 34/10 & 340 & 39.29 & 22 & 0.0122 & 0.5065 \\
\texttt{1.1.1.1.1.2.1.1} & 8 & 34 & 27/7 & 189 & 36.85 & 20 & 0.0055 & 0.5143 \\
\texttt{1.1.1.1.1.2.1.1.1} & 9 & 27 & 16/11 & 176 & 15.89 & 21 & 0.0496 & 0.3417 \\
\texttt{1.1.1.1.1.2.1.1.1.1} & 10 & 16 & 14/2 & 28 & 11.34 & 14 & 0.0259 & 0.4893 \\
\texttt{1.1.1.2} & 4 & 19 & 10/9 & 90 & 12.73 & 9 & 0.0614 & 0.3960 \\
\texttt{1.2} & 2 & 42 & 27/15 & 405 & 17.19 & 16 & 0.0421 & 0.3808 \\
\texttt{1.2.1} & 3 & 27 & 18/9 & 162 & 10.85 & 12 & 0.0738 & 0.2612 \\
\texttt{1.2.1.1} & 4 & 18 & 9/9 & 81 & 23.42 & 20 & 0.0426 & 0.3889 \\
\texttt{1.2.2} & 3 & 15 & 10/5 & 50 & 11.35 & 9 & 0.0848 & 0.4137 \\
\bottomrule
\end{tabular}
\caption{All 28 surgeries performed by ARCUS on the Nikkei-225 ($\alpha=0.5$, $G_{\min}=10$, $N_{\min}=15$). Node identifiers encode the path from the root (\texttt{.1} = part A, \texttt{.2} = part B); Level = depth + 1. $[\kappa_{\text{lo}},\kappa_{\text{hi}}]$ is the largest curvature gap at the stopping iteration. All 28 nodes stopped by the $G$-and-stability criterion. Children with $N<15$ are terminal and have no row.}
\label{tab:full}
\end{table}

%% file: tables/tab_members.tex
\begingroup\footnotesize
\begin{longtable}{l l r r r r p{8.3cm}}
\caption{Membership of the 29 ARCUS terminal clusters (identifiers as in Figure~\ref{fig:hier}), with Tokyo Stock Exchange codes; together the clusters list all 226 nodes. $\rho_{\text{in}}$: mean pairwise correlation within the cluster; $\rho_{\text{out}}$: mean correlation between its members and all other stocks; $p$: share of 2{,}000 random groups, of the same size as the cluster without placeholders, with a larger $\rho_{\text{in}}-\rho_{\text{out}}$ (for C28, groups of 8). Correlation statistics exclude the two placeholder nodes. $^\dagger$: no price data, zero-filled placeholder. $^\ast$: listed during the window, so its correlations use a shorter sample (Section~\ref{sec:data}).}\label{tab:members}\\
\toprule
ID & Node & $N$ & $\rho_{\text{in}}$ & $\rho_{\text{out}}$ & $p$ & Members \\
\midrule
\endfirsthead
\toprule
ID & Node & $N$ & $\rho_{\text{in}}$ & $\rho_{\text{out}}$ & $p$ & Members \\
\midrule
\endhead
C1 & \texttt{\scriptsize 1.1.1.1.1.1.1.1.1.1.1.1.1.1} & 14 & 0.55 & 0.41 & $<$0.001 & Credit Saison (8253); Daiwa Securities Group (8601); Dowa (5714); IHI (7013); JGC (1963); Kawasaki Heavy Industries (7012); Mitsubishi Heavy Industries (7011); Mitsubishi Materials (5711); Nomura (8604); Orix (8591); Sojitz (2768); Sumitomo Heavy Industries (6302); Tokyo Tatemono (8804); Toyota Tsusho (8015) \\
C2 & \texttt{\scriptsize 1.1.1.1.1.1.1.1.1.1.1.1.1.2} & 3 & 0.79 & 0.39 & $<$0.001 & JFE (5411); Kobe Steel (5406); Nippon Steel (5401) \\
C3 & \texttt{\scriptsize 1.1.1.1.1.1.1.1.1.1.1.1.2} & 3 & 0.70 & 0.35 & $<$0.001 & ENEOS (5020); Idemitsu Kosan (5019); Inpex (1605) \\
C4 & \texttt{\scriptsize 1.1.1.1.1.1.1.1.1.1.1.2} & 5 & 0.77 & 0.44 & $<$0.001 & Itochu (8001); Marubeni (8002); Mitsubishi (8058); Mitsui \& Co. (8031); Sumitomo (8053) \\
C5 & \texttt{\scriptsize 1.1.1.1.1.1.1.1.1.1.2.1} & 9 & 0.43 & 0.37 & 0.028 & Aozora Bank (8304); Denka (4061); Japan Post (6178); Kuraray (3405); Nippon Express (9147)$^\ast$; Oji (3861); TOPPAN (7911); Taiheiyo Cement (5233); Teijin (3401) \\
C6 & \texttt{\scriptsize 1.1.1.1.1.1.1.1.1.1.2.2} & 7 & 0.61 & 0.42 & $<$0.001 & Asahi Kasei (3407); Mitsubishi Chemical (4188); Mitsui Chemicals (4183); Sumitomo Chemical (4005); Toray Industries (3402); Tosoh (4042); UBE (4208) \\
C7 & \texttt{\scriptsize 1.1.1.1.1.1.1.1.1.2} & 5 & 0.70 & 0.40 & $<$0.001 & Dai-ichi Life Insurance (8750); MS\&AD Insurance Group (8725); Sompo (8630); T\&D (8795); Tokio Marine (8766) \\
C8 & \texttt{\scriptsize 1.1.1.1.1.1.1.1.2} & 3 & 0.85 & 0.29 & $<$0.001 & Kawasaki Kisen Kaisha (9107); Mitsui O.S.K. Lines (9104); Nippon Yusen (9101) \\
C9 & \texttt{\scriptsize 1.1.1.1.1.1.1.2.1.1} & 10 & 0.53 & 0.40 & $<$0.001 & Canon (7751); Dentsu (4324); JTEKT (6473); Japan Airlines (9201); Konica Minolta (4902); NSK (6471); NTN (6472); Ricoh (7752); Seiko Epson (6724); Tokyu Fudosan (3289) \\
C10 & \texttt{\scriptsize 1.1.1.1.1.1.1.2.1.2} & 11 & 0.61 & 0.40 & $<$0.001 & Bridgestone (5108); Honda Motor (7267); Isuzu Motors (7202); Mazda Motor (7261); Mitsubishi Motors (7211); Nissan Motor (7201); Subaru (7270); Suzuki Motor (7269); Toyota Motor (7203); Yamaha Motor (7272); Yokohama Rubber (5101) \\
C11 & \texttt{\scriptsize 1.1.1.1.1.1.1.2.2} & 3 & 0.77 & 0.34 & $<$0.001 & Isetan Mitsukoshi (3099); J. Front Retailing (3086); Takashimaya (8233) \\
C12 & \texttt{\scriptsize 1.1.1.1.1.1.2.1.1} & 11 & 0.36 & 0.33 & 0.123 & Asahi Group (2502); Comsys (1721); Dai Nippon Printing (7912); Japan Tobacco (2914); Kansai Electric Power (9503); Nissui (1332); Osaka Gas (9532); Ryohin Keikaku (7453); Secom (9735); Seven \& I (3382); Tokyo Electric Power (9501) \\
C13 & \texttt{\scriptsize 1.1.1.1.1.1.2.1.2} & 4 & 0.43 & 0.31 & 0.007 & Astellas Pharma (4503); Otsuka (4578); Sumitomo Pharma (4506); Takeda Pharmaceutical (4502) \\
C14 & \texttt{\scriptsize 1.1.1.1.1.1.2.2} & 10 & 0.60 & 0.40 & $<$0.001 & Daiwa House Industry (1925); Haseko (1808); Kajima (1812); Mitsubishi Estate (8802); Mitsui Fudosan (8801); Obayashi (1802); Sekisui House (1928); Shimizu (1803); Sumitomo Realty \& Development (8830); Taisei (1801) \\
C15 & \texttt{\scriptsize 1.1.1.1.1.2.1.1.1.1.1} & 14 & 0.47 & 0.39 & 0.001 & AGC (5201); Amada (6113); Casio Computer (6952); Denso (6902); GS Yuasa (6674); Kanadevia (7004); Kubota (6326); Mitsubishi Electric (6503); Nikon (7731); Panasonic (6752); Sumitomo Electric Industries (5802); Sumitomo Metal Mining (5713); Tokai Carbon (5301); Tokuyama (4043) \\
C16 & \texttt{\scriptsize 1.1.1.1.1.2.1.1.1.1.2} & 2 & 0.73 & 0.43 & 0.004 & Hitachi Construction Machinery (6305); Komatsu (6301) \\
C17 & \texttt{\scriptsize 1.1.1.1.1.2.1.1.1.2} & 11 & 0.55 & 0.41 & $<$0.001 & Alps Alpine (6770); FANUC (6954); Kyocera (6971); MinebeaMitsumi (6479); NGK Insulators (5333); Nippon Electric Glass (5214); Nissan Chemical (4021); Nitto Denko (6988); Okuma (6103); Yamaha (7951); Yokogawa Electric (6841) \\
C18 & \texttt{\scriptsize 1.1.1.1.1.2.1.1.2} & 7 & 0.47 & 0.36 & 0.001 & Ebara (6361); Fuji Electric (6504); Hitachi (6501); Japan Steel Works (5631); Resonac (4004); Socionext (6526)$^\ast$; SoftBank Group (9984) \\
C19 & \texttt{\scriptsize 1.1.1.1.1.2.1.2} & 10 & 0.42 & 0.35 & 0.011 & Daikin Industries (6367); Fujifilm (4901); Japan Exchange Group (8697); Omron (6645); Rakuten (4755); SMC (6273); Sharp (6753); Shin-Etsu Chemical (4063); Sony (6758); Toto (5332) \\
C20 & \texttt{\scriptsize 1.1.1.1.1.2.2} & 3 & 0.55 & 0.33 & 0.002 & Fujikura (5803); Furukawa Electric (5801); Mitsui Mining \& Smelting (5706) \\
C21 & \texttt{\scriptsize 1.1.1.1.2} & 9 & 0.82 & 0.40 & $<$0.001 & Chiba Bank (8331); Concordia Financial Group (7186); Fukuoka Financial Group (8354); Mitsubishi UFJ Financial Group (8306); Mizuho Financial Group (8411); Resona (8308); Shizuoka Financial Group (5831)$^\ast$; Sumitomo Mitsui Financial Group (8316); Sumitomo Mitsui Trust (8309) \\
C22 & \texttt{\scriptsize 1.1.1.2.1} & 10 & 0.63 & 0.32 & $<$0.001 & ANA (9202); Central Japan Railway (9022); East Japan Railway (9020); Keio (9008); Keisei Electric Railway (9009); Marui Group (8252); Odakyu Electric Railway (9007); Tobu Railway (9001); Tokyu (9005); West Japan Railway (9021) \\
C23 & \texttt{\scriptsize 1.1.1.2.2} & 9 & 0.35 & 0.27 & 0.008 & Chubu Electric Power (9502); KDDI (9433); Kirin (2503); NH Foods (2282); Nippon Telegraph \& Telephone (9432); Nisshin Seifun Group (2002); Sapporo (2501); SoftBank (9434); Tokyo Gas (9531) \\
C24 & \texttt{\scriptsize 1.1.2} & 11 & 0.57 & 0.33 & $<$0.001 & Advantest (6857); Disco (6146); Murata Manufacturing (6981); Renesas Electronics (6723); Rohm (6963); SCREEN (7735); SUMCO (3436); TDK (6762); Taiyo Yuden (6976); Tokyo Electron (8035); Yaskawa Electric (6506) \\
C25 & \texttt{\scriptsize 1.2.1.1.1} & 9 & 0.34 & 0.26 & 0.005 & Bandai Namco (7832); DeNA (2432); Fujitsu (6702); Konami (9766); NEC (6701); Nexon (3659); Nintendo (7974); Nomura Research Institute (4307); Trend Micro (4704) \\
C26 & \texttt{\scriptsize 1.2.1.1.2} & 9 & 0.38 & 0.31 & 0.018 & Fast Retailing (9983); Hoya (7741); Keyence (6861); Lasertec (6920); Nidec (6594); Olympus (7733); Recruit (6098); Shiseido (4911); Terumo (4543) \\
C27 & \texttt{\scriptsize 1.2.1.2} & 9 & 0.32 & 0.19 & $<$0.001 & Baycurrent (6532); CyberAgent (4751); LY (4689); M3 (2413); Mercari (4385); Nitori (9843); SHIFT (3697); Yamato (9064); ZOZO (3092) \\
C28 & \texttt{\scriptsize 1.2.2.1} & 10 & 0.31 & 0.25 & 0.049 & Aeon (8267); Ajinomoto (2802); Hino Motors (7205)$^\dagger$; Kao (4452); Kikkoman (2801); Meiji (2269); NTT Data (9613)$^\dagger$; Nichirei (2871); Oriental Land (4661); Toho (9602) \\
C29 & \texttt{\scriptsize 1.2.2.2} & 5 & 0.36 & 0.25 & 0.006 & Chugai Pharmaceutical (4519); Daiichi Sankyo (4568); Eisai (4523); Kyowa Kirin (4151); Shionogi (4507) \\
\bottomrule
\end{longtable}
\endgroup